\documentclass[11pt]{article}% добавляется twoside для двусторонней печати

\usepackage[cp1251]{inputenc}
\usepackage[T1]{fontenc}
\usepackage{textcomp}
\usepackage[centertags]{amsmath}
\usepackage{amsfonts}
\usepackage{amssymb}
\usepackage[numbers,sort&compress]{natbib}% упорядочивает ссылки
\usepackage{graphicx}
\usepackage[hypertex]{hyperref}%hypertex

\usepackage{paperinitial}% Установка параметров страницы

\paperinitialization{15mm}{15mm}{15mm}{15mm}{2pt}{10pt}

\newcommand{\lan}{\langle}
\newcommand{\ran}{\rangle}

\newcommand{\vf}{\varphi}
\newcommand{\vk}{\varkappa}

\newcommand{\al}{\alpha}
\newcommand{\be}{\beta}
\newcommand{\ga}{\gamma}

\newcommand{\de}{\delta}

\newcommand{\la}{\lambda}

\newcommand{\spx}{\mathbf{x}}

\begin{document}
\allowdisplaybreaks[4]% позволяет переносить многострочные формулы
\frenchspacing% уменьшение пробелов после запятых
\setlength{\unitlength}{1pt}% устанавливает единицу длины в окружении picture
%\selectlanguage{english}

\title{{\Large\textbf{Radiation of twisted photons in elliptical undulators}}}

\date{}

\author{P.O. Kazinski\thanks{E-mail: \texttt{kpo@phys.tsu.ru}}\;\, and V.A. Ryakin\thanks{E-mail: \texttt{vlad-r@sibmail.com}}\\[0.5em]
{\normalsize Physics Faculty, Tomsk State University, Tomsk 634050, Russia}
}

\maketitle

\begin{abstract}

The explicit expressions for the average number of twisted photons radiated by a charged particle in an elliptical undulator in the classical approximation as well as in the approach accounting for the quantum recoil are obtained. It is shown that radiation emitted by a particle moving along an elliptical helix which evolves around the axis specifying the angular momentum of twisted photons obeys the selection rule: $m+n$ is an even number, where $m$ is a projection of the total angular momentum of a twisted photon and $n$ is the harmonic number of the undulator radiation. This selection rule is a generalization of the previously known selection rules for radiation of twisted photons by circular and planar undulators and it holds for both classical and quantum approaches. The class of trajectories of charged particles that produce the twisted photon radiation obeying the aforementioned selection rule is described.

\end{abstract}

\section{Introduction}

Nowadays, the undulators are the standard tool for generating a powerful electromagnetic radiation with specified properties across a wide range of energies of radiated photons. It is known that the helical and planar undulators can be used as a bright source of twisted photons \cite{SasMcNu,AfanMikh,BordKN,HemMar12,BHKMSS,HKDXMHR,RibGauNin14}. The properties of such photons were thoroughly investigated in \cite{SasMcNu,BKL2,BKL4,BKL6}. In this paper, we generalize the previously known results to the case of an elliptical undulator where the charged particles move along elliptical helices. To our knowledge, the properties of twisted photons generated by such devices in the non-dipole regime have not been described until now. We will derive the explicit expressions for the average number of twisted photons produced by an elliptical undulator with account for the quantum recoil experienced by the charged particle in emitting a hard photon. In particular, we shall prove the selection rule that $m+n$ is an even number, where $m$ is the projection of the total angular momentum onto the undulator axis and $n$ is the number of the undulator radiation harmonic. This rule is in agreement with the selection rules obtained for the twisted photons emitted by circular and planar undulators \cite{BKL2,BKL4,BKL6}.

The twisted photons are the quanta of the electromagnetic field with the definite helicity $s$, the projection $m$ of the total angular momentum on the propagation axis of the photon, the momentum projection on the same axis, and the modulus of the perpendicular momentum component $k_\perp$ \cite{JenSerprl,JenSerepj}. These stationary states form a complete set, they are the eigenvectors of the projection of total angular momentum operator on the axis $3$ and, in the paraxial approximation, $k_\perp/|k_3|\ll1$, they possess the projection $l=m-s$ of the orbital angular momentum on this axis. These states of the electromagnetic field are used in optical tweezers; in microscopy to increase a contrast of the pictures; in telecommunication and quantum cryptography where the projection of the angular momentum is used as an additional quantum number carrying information; in the studies of rotational degrees of freedom of quantum objects where the twisted photons stimulate non-dipole transitions (see, for review, \cite{TorTorTw,AndBabAML,PadgOAM25,Roadmap16,KnyzSerb}). Moreover, the investigations of the radiation properties in the basis of twisted photons allow one to establish the properties that are difficult to unveil in the basis of plane-wave photons, which is commonly used to describe radiation. Currently, there are plenty of designs for detectors enabling one to register twisted photons, to find the number of them and their quantum numbers in the incident radiation \cite{aperture,BLCBP,SSDFGCY,LavCourPad,Walsh,RGMMSCFR,RGMMSCFR2,PSP20,GoTsuKu,taira2019}. Therefore, the expressions for the average number of twisted photons that are derived in the present paper can be directly verified in experiments.

The paper is organized as follows. In Sec. \ref{Gen_Form}, for the reader convenience, the general formulas that are used to describe the radiation of twisted photons in elliptical undulators are given. Section \ref{Rad_Tw_Phot} is the main part of the paper. It contains the explicit formulas for the average number of radiated twisted photons as well as their analysis. In Conclusion, we recapitulate the results and formulate the possible directions for further research. Appendix \ref{Sel_Rul} contains the proof of the selection rule. Some properties of the special functions appearing in the formula for the average number of radiated twisted photons are collected in Appendix \ref{Gnm_Prop}. The notation introduced in \cite{BKL2,BKL4,BKL6} is vastly used. In particular, we use the system of units such that $\hbar=c=1$ and the fine structure constant $\al=e^2/(4\pi)$. For consistency of notation, it is always supposed that $x\equiv x_1$, $y\equiv x_2$, and $z\equiv x_3$.

%\newpage
\section{General formulas}\label{Gen_Form}

In this section, we provide some general formulas that are necessary for evaluation of the average number of twisted photons radiated by a charged particle in elliptical undulators. The trajectory of the particle with the charge $e$ in the elliptical undulator with the section length $\la_0$ takes form (see, e.g., \cite{Bord.1}, Ch. 5)
\begin{equation}\label{trajectory}
    x=x_0+b_x\cos\vf,\qquad y=y_0+b_y\sin\vf,\qquad z=z_0+\be_3 t+b_3\sin(2\vf),
\end{equation}
where $t$ is the laboratory time, $\vf=\omega t-\chi$, the parameters $x_0$, $y_0$, $z_0$, $\chi$ are some constants, $\be_3$ is the average particle velocity along the $z$ axis, and $\omega=2\pi\be_3/\la_0>0$. The amplitudes of the particle oscillations along different axes are expressed via the magnetic field strength in the undulator
\begin{equation}\label{axyz}
    b_{x}=\frac{\la_0^2 H_{y}}{4\pi^2\ga},\qquad b_{y}=-\frac{\la_0^2 H_{x}}{4\pi^2\ga},\qquad b_3=\frac{\la_0^3(H_y^2-H_x^2)}{64\pi^3\ga^2},
\end{equation}
where $\ga$ is the Lorentz-factor of the particle, the magnetic field strength $\mathbf{H}$ is measured in the units of the critical field
\begin{equation}\label{crit_field}
    H_0=m^2_e/|e|\approx 4.41\times 10^{13}\;\text{G},
\end{equation}
and the lengths are measured in the units of the Compton wavelength, $1/m_e\approx3.86\times 10^{-11}$ cm. The undulator strength parameter for the trajectory \eqref{trajectory} is written as
\begin{equation}
    K:=\ga\lan\be_\perp^2\ran^{1/2}=\la_0\frac{\sqrt{H_x^2+H_y^2}}{2\sqrt{2}\pi}.
\end{equation}
By the order of magnitude, the oscillation amplitudes are estimated as
\begin{equation}\label{estimations}
    b_{x,y}^2\approx \frac{K^2}{\omega^2\ga^2},\qquad |b_3|\approx\frac{K^2}{2\pi \omega\ga^2}.
\end{equation}
The expression \eqref{trajectory} is obtained under the assumption that $\ga\gg1$ and $K/\ga\ll1$. If the condition $K\ll1$ is satisfied, the undulator is in the dipole regime which was thoroughly investigated in \cite{BKL2,BKL4}. Henceforth we assume that $K\gtrsim1$ and so the dipole approximation is inapplicable.

The trajectory of the charged particle takes the form \eqref{trajectory} for $t\in[-TN/2,TN/2]$, where $T:=2\pi/\omega$ and $N\gg1$ is the number of undulator sections. If $t$ does not belong to the aforementioned interval, we suppose that the particle moves along the $z$ axis with the velocity $\be_\parallel=\sqrt{1-1/\ga^2}$. We will investigate the properties of radiation from the part of the trajectory $t\in[-TN/2,TN/2]$. The radiation formed on this interval dominates when $N\gg1$ for the photon energies corresponding to the harmonics of undulator radiation.

The average number of twisted photons emitted by a classical point charge is given by \cite{BKL2}
\begin{multline}\label{probabil}
    dP(s,m,k_3,k_\perp)=e^2\bigg|\int d\tau e^{-i[k_0x^0(\tau)-k_3x_3(\tau)]}\Big\{\frac12\big[\dot{x}_+(\tau)a_-(s,m,k_3,k_\perp;\spx(\tau))+\\
    +\dot{x}_-(\tau)a_+(s,m,k_3,k_\perp;\spx(\tau)) \big]
    +\dot{x}_3(\tau)a_3(m,k_\perp;\spx(\tau))\Big\} \bigg|^2 n_\perp^3\frac{dk_3dk_\perp}{16\pi^2},
\end{multline}
where $s$ is the twisted photon helicity, $m$ is the projection of the total angular momentum onto the $z$ axis, $k_\perp$ and $k_3$ are the projections of the photon momentum, $k_0=\sqrt{k_3^2+k_\perp^2}$ is the photon energy, $n_\perp:=k_\perp/k_0$ (see, for more details, \cite{JenSerprl,JenSerepj,BKL2,BKL4,BKL6}). We also use the following notation
\begin{equation}
    x_\pm:=x\pm iy.
\end{equation}
The explicit expressions for the mode functions of twisted photons $a_\pm$, $a_3$ are given, e.g., in formula (2.5) of the paper \cite{BKL6}. The parameter $\tau$ in \eqref{probabil} is an arbitrary parameter on the particle worldline. In our case, it is convenient to choose it as $\tau=t\equiv x^0(t)$.

\section{Radiation of twisted photons}\label{Rad_Tw_Phot}

In order to find the average number of radiated twisted photons, one has to evaluate the integrals in \eqref{probabil}. Let us introduce the notation
\begin{equation}\label{I_integrals}
\begin{split}
    I_3&:=\int_{-TN/2}^{TN/2}dt e^{-ik_0[t -n_3(z_0+\be_3 t+b_3\sin(2\vf))]}\dot{x}_3a_3(m,k_\perp;\spx(t)),\\
    I_\pm&:=\int_{-TN/2}^{TN/2}dt e^{-ik_0[t -n_3(z_0+\be_3 t+b_3\sin(2\vf))]}\dot{x}_\pm a_\mp(s,m,k_3,k_\perp;\spx(t)),
\end{split}
\end{equation}
where $n_3:=k_3/k_0$. As we shall see, for $N\gg1$, these integrals give the main contribution to \eqref{probabil}, i.e., the contributions of the edge radiation to \eqref{probabil} are negligible in this case. Then
\begin{equation}\label{probab_I}
    dP(s,m,k_3,k_\perp)\approx e^2\big|I_3+(I_++I_-)/2\big|^2 n_\perp^3\frac{dk_3dk_\perp}{16\pi^2}.
\end{equation}
The components of the particle trajectory \eqref{trajectory} are written as
\begin{equation}
    x_\pm=x_\pm^0+Re^{\pm i\vf}+ D e^{\mp i\vf},
\end{equation}
where
\begin{equation}
    R:=(b_x+b_y)/2,\qquad D:=(b_x-b_y)/2,\qquad x_\pm^0:=x_0\pm i y_0.
\end{equation}
The components of the velocity become
\begin{equation}\label{dotxpm}
    \dot{x}_\pm=\pm i\omega(R e^{\pm i\vf}- D e^{\mp i\vf}),\qquad \dot{x}_3=\be_3+2\omega b_3\cos(2\vf).
\end{equation}
The contribution of the second term in the last expression is negligible when $n_\perp\ga\lesssim\max(1,K)$, i.e., in the domain of parameters where the main part of radiation is concentrated.

Let us start with the integral $I_3$. It is convenient to use the addition theorem for the Bessel functions (see Eq. (A6) of \cite{BKL2}) in the form
\begin{equation}\label{add_thm}
    j_\nu(x_++y_+,x_-+y_-)=\sum_{n=-\infty}^\infty j_{\nu-n}(x_+,x_-)j_n(y_+,y_-),
\end{equation}
and the property
\begin{equation}\label{j_prop}
    j_m(ap,q/a)=a^m j_m(p,q),\qquad m\in \mathbb{Z}.
\end{equation}
The definition of the functions $j_\nu(p,q)$ and their properties are given in Appendix A of \cite{BKL2}. Employing these expressions, we obtain
\begin{equation}\label{h_3}
    j_{m}(k_\perp x_+,k_\perp x_-)e^{ik_3b_3\sin(2\vf)}=\sum_{n,k,r=-\infty}^\infty j_{m-n-2k+2r}^0J_{n+k-2r}(\rho)J_k(\de)J_r(\vk)e^{in\vf},
\end{equation}
where the exponent on the left-hand side of the expression is expanded by using the Jacobi-Anger identity. Besides, the shorthand notation has been introduced
\begin{equation}
    j^0_m:=j_m(k_\perp x_+^0,k_\perp x_-^0),\qquad \rho:=k_\perp R,\qquad \de:=k_\perp D,\qquad\vk=k_3b_3.
\end{equation}
The expression \eqref{h_3} depends on $t$ only through the exponent standing on the right-hand side. Hence, the integral over $t$ is readily evaluated,
\begin{equation}\label{I_3}
    I_3=2\pi\be_3 \sum_{n,k,r=-\infty}^\infty \de_N\big(k_0(1-n_3\be_3)-n\omega\big)e^{ik_3z_0-in\chi}j_{m-n-2k+2r}^0J_{n+k-2r}(\rho)J_k(\de)J_r(\vk),
\end{equation}
where
\begin{equation}
    \de_N(x):=\frac{\sin(TNx/2)}{\pi x}.
\end{equation}
The integrals $ I_\pm $ are evaluated in a similar way. In this case, we obtain
\begin{equation}\label{I_pm}
\begin{split}
    I_\pm =& - 2\pi\omega\frac{n_3\pm s}{n_\perp} \sum_{n,k,r=-\infty}^\infty \de_N\big(k_0(1-n_3\be_3)-n\omega\big) e^{ik_3z_0-in\chi} j_{m-n-2k+2r}^0 \times\\
    &\times\big[RJ_{n+k-2r\mp1}(\rho) J_k(\de) - D J_{n+k-2r}(\rho) J_{k\mp1}(\de)\big] J_r(\vk).
\end{split}
\end{equation}
The two terms in the square brackets come from the two terms in the expression for $\dot{x}_\pm$ in \eqref{dotxpm}. For $N$ large, the expressions \eqref{I_3}, \eqref{I_pm} possess the sharp maxima at
\begin{equation}
    k_0=\frac{n\omega}{1-n_3\be_3},\qquad n=\overline{1,\infty},
\end{equation}
that correspond to the undulator radiation harmonics numerated by $n$.

Thus, the one-particle amplitude of twisted photon radiation is proportional to
\begin{equation}
\begin{split}
    I_3+(I_++I_-)/2=& - 2\pi \sum_{n,k,r=-\infty}^\infty \de_N\big(k_0(1-n_3\be_3)-n\omega\big) e^{ik_3z_0-in\chi} j_{m-n-2k+2r}^0 J_r(\vk)\times\\
    &\times \Big\{\Big(\frac{\omega(n-2r)n_3}{k_\perp n_\perp}-\be_3 \Big)J_{n+k-2r}(\rho)J_k(\de)+\\
    &+\frac{s\omega}{n_\perp}\big[RJ'_{n+k-2r}(\rho)J_k(\de) -D J_{n+k-2r}(\rho)J'_k(\de)\big] \Big\}.
\end{split}
\end{equation}
This expression for the radiation amplitude can be used for description of the both coherent and incoherent radiations of twisted photons by particle beams \cite{BKb,BKL5,BKLb}. It is clear now that the second term in $\dot{x}_3$ in Eq. \eqref{dotxpm} substituted into \eqref{probab_I} gives rise to the correction to \eqref{probab_I} that is of the order or less than $n_\perp^2$ in comparison with the other terms in this expression. Therefore, this contribution can be discarded for $n_\perp^2\ll1$.

Neglecting the interference terms between different harmonics in \eqref{probab_I}, we find the average number of twisted photons \eqref{probabil} emitted by a charged particle in an elliptic undulator
\begin{equation}\label{probab_ell1}
\begin{split}
    dP(s,m,k_3,k_\perp)\approx&\,\frac{e^2}{4}\sum_{n=1}^\infty \de^2_N\big(k_0(1-n_3\be_3)-n\omega\big) \bigg|\sum_{k,r=-\infty}^\infty j_{m-n-2k+2r}^0 J_r(\vk)\times\\
    &\times \Big\{\Big(\frac{\omega(n-2r)n_3}{k_\perp n_\perp}-\be_3 \Big)J_{n+k-2r}(\rho)J_k(\de)+\\
    &+\frac{s\omega}{n_\perp}\big[RJ'_{n+k-2r}(\rho)J_k(\de) -D J_{n+k-2r}(\rho)J'_k(\de)\big] \Big\} \bigg|^2 n_\perp^3dk_3dk_\perp.
\end{split}
\end{equation}
In the case $k_\perp|x^0_+|\ll1$, i.e., when the center line of the elliptical helix trajectory is close to the axis used to define the angular momentum of twisted photons, the following relation is valid
\begin{equation}\label{center}
    j^0_{m-n-2k+2r}\approx\de_{m,n+2k-2r}.
\end{equation}
Therefore, the radiation of twisted photons obeys the selection rule: $m+n$ must be an even number. The same selection rule holds for radiation of twisted photons by charged particles moving in a planar undulator \cite{BKL2,BKL4,BKL6}. The circular undulator radiation obeys the selection rule $m=\pm n$, the sign being specified by the helix chirality \cite{SasMcNu,BKL2,BKL4,KatohPRL,KatohSRexp,EppGusel19}. It is evident that $m+n$ is an even number in this case too. It should be noted that accounting for the second term in the expression for $\dot{x}_3$ in \eqref{dotxpm} does not spoil this selection rule. In general, such a selection rule is valid for the twisted photons radiated by the charged particle moving along the trajectory
\begin{equation}\label{traj_odd}
    x_+=\sum_{k=-\infty}^\infty c_k e^{i(2k+1)\vf},\qquad x_-=\sum_{k=-\infty}^\infty c^*_k e^{-i(2k+1)\vf},\qquad z=\be_3 t+\sum_{k=-\infty}^\infty d_k e^{2ik\vf},
\end{equation}
where $d_k^*=d_{-k}$. The proof of this statement is postponed to Appendix \ref{Sel_Rul}.

Using the relation \eqref{center}, the expression \eqref{probab_ell1} is simplified to
\begin{equation}\label{probab_ell2}
\begin{split}
    \frac{dP(s,m,k_3,k_\perp)}{n_\perp^3dk_3 dk_\perp}\approx&\,\frac{e^2}{4} \sum_{n=1}^\infty \de^2_N\big(k_0(1-n_3\be_3)-n\omega\big) \bigg[\sum_{r=-\infty}^\infty J_r(\vk)\times\\
    &\times \Big\{\Big(\frac{\omega(n-2r)n_3}{k_\perp n_\perp}-\be_3 \Big)J_{(m+n)/2-r}(\rho)J_{(m-n)/2+r}(\de)+\\
    &+\frac{s\omega}{n_\perp}\big[RJ'_{(m+n)/2-r}(\rho)J_{(m-n)/2+r}(\de) -D J_{(m+n)/2-r}(\rho)J'_{(m-n)/2+r}(\de)\big] \Big\} \bigg]^2.
\end{split}
\end{equation}
One can see that the sign flip of the elliptical helix chirality changes the average number of twisted photons as follows
\begin{equation}\label{dP_transform}
    dP(s,m,k_\perp,k_3)\rightarrow dP(-s,-m,k_\perp,k_3).
\end{equation}
This transformation becomes a symmetry for a planar undulator \cite{BKL2,BKL4,BKL6,BKL8}. In the cases of planar and circular undulators, formula \eqref{probab_ell2} reproduces exactly the expressions for the average number of twisted photons derived in \cite{BKL2}.

Now we take into account the quantum recoil effect. We will employ the method developed in \cite{BKL4,BKL6}. This method is an adaptation of the Baier-Katkov method \cite{BaiKat1,BaKaStrbook} for description of the quantum radiation produced by charged ultrarelativistic scalar and Dirac particles in the domain of parameters where the main part of radiation is concentrated. As long as $P_0:=m_e\ga$ is an integral of motion of a charged particle in an undulator, we can use formula (2.13) of \cite{BKL6} to describe the radiation of twisted photons. Let
\begin{equation}
    q:=P_0/(P_0-k_0),
\end{equation}
and
\begin{equation}\label{kop}
    k_0':=q[k_0-k_\perp^2/(2 P_0)],\qquad k_3'=qk_3,\qquad\vk':=qk_3 b_3.
\end{equation}
Employing formula (2.13) of \cite{BKL6}, we come to the integrals of the form \eqref{I_integrals} with the obvious replacements. We denote these integrals as $I'_3$ and $I'_\pm$. Then, disregarding the irrelevant common phase, we have
\begin{equation}
\begin{split}
    I'_3=&\, 2\pi \sum_{n,k,r=-\infty}^\infty \de_N\big(k'_0-k'_3\be_3-n\omega\big)e^{-in\chi}j_{m-n-2k+2r}^0J_{n+k-2r}(\rho)J_k(\de)J_r(\vk'),\\
    I'_s=&-\frac{4\pi\omega}{n_\perp} \sum_{n,k,r=-\infty}^\infty \de_N\big(k'_0-k'_3\be_3-n\omega\big) e^{-in\chi} j_{m-n-2k+2r}^0 \times\\
    &\times\big[RJ_{n+k-2r-s}(\rho) J_k(\de) - D J_{n+k-2r}(\rho) J_{k-s}(\de)\big] J_r(\vk'),
\end{split}
\end{equation}
where we have taken into account that $n_3\approx1$ within the approximations made.

The energy of photons radiated at the $n$-th harmonic with account for the quantum recoil is deduced by setting the argument of $\de_N(x)$ to zero. If we neglect the second term in the square brackets in the expression for $k_0$ in \eqref{kop}, which is small within the approximation considered, then we find the following spectrum of undulator radiation
\begin{equation}
    k_0=\frac{n\omega}{1-n_3\be_3+n\omega/P_0},\qquad n=\overline{1,\infty}.
\end{equation}
Neglecting the interference between different harmonics, we obtain the inclusive probability of radiation of one twisted photon by a charged Dirac particle in an elliptical undulator in the leading order of perturbation theory
\begin{equation}\label{probab_recoil1}
\begin{split}
    \frac{dP(s,m,k_3,k_\perp)}{n_\perp^3dk_3dk_\perp}\approx&\,\frac{e^2}{4}\frac{1+q^2}{2}\sum_{n=1}^\infty \de^2_N\big(k'_0-k'_3\be_3-n\omega\big) \bigg|\sum_{k,r=-\infty}^\infty j_{m-n-2k+2r}^0 J_r(\vk')\times\\
    &\times \Big\{ \big[J_{n+k-2r}(\rho) -\frac{\omega R}{n_\perp} J_{n+k-2r-s}(\rho)\big] J_k(\de)+ \frac{\omega D}{n_\perp} J_{n+k-2r}(\rho)J_{k-s}(\de) \Big\} \bigg|^2 .
\end{split}
\end{equation}
For $k_\perp|x_+^0|\ll1$, the relation \eqref{center} holds and so does the selection rule $m+n$ is an even number, i.e., this selection rule is valid in the quantum case as well.

When the condition $k_\perp|x_+^0|\ll1$ is met, the expression \eqref{probab_recoil1} can be simplified. Let us introduce the functions
\begin{equation}\label{gmn}
    g_{nm}(\vk,\rho,\de):=\frac{1+(-1)^{n+m}}{2}\sum_{r=-\infty}^\infty J_r(\vk)J_{(m+n)/2-r}(\rho)J_{(m-n)/2+r}(\de).
\end{equation}
In fact, these functions are the coefficients of the Fourier series \eqref{h_3} when $x^0_\pm=0$. Some properties of the functions \eqref{gmn} are presented in Appendix \ref{Gnm_Prop}. Then the probability \eqref{probab_recoil1} takes the form
\begin{equation}\label{probab_recoil2}
    \frac{dP(s,m,k_3,k_\perp)}{n_\perp^3dk_3dk_\perp}=\frac{e^2}{4}\frac{1+q^2}{2}\sum_{n=1}^\infty \de^2_N\big(k'_0-k'_3\be_3-n\omega\big)\Big(g_{nm} -\frac{\omega R}{n_\perp} g_{n-s,m-s} + \frac{\omega D}{n_\perp} g_{n+s,m-s}\Big)^2,
\end{equation}
where $g_{nm}\equiv g_{nm}(\vk',\rho,\de)$. In order to find the probability of radiation of twisted photons by scalar particles, one should replace the common factor $(1+q^2)/2$ by $q$ (see (2.10) in \cite{BKL6}). In the particular cases of circular and planar undulators, formula \eqref{probab_recoil2} reproduces formulas (54) of \cite{BKL4} and (3.7) of \cite{BKL6}. Furthermore, assuming that the quantum recoil is small and employing the paraxial approximation $n^2_\perp\ll1$, formula \eqref{probab_recoil2} turns into the expression \eqref{probab_ell2}. Using the symmetry property \eqref{g_prop}, it is not difficult to show that the sign flip of the chirality of the elliptic helix trajectory of the particle in the undulator results in the transformation rule \eqref{dP_transform} for the probability of radiation of twisted photons \eqref{probab_recoil2}.

%\newpage
\section{Conclusion}\label{Conclus}

Let us briefly summarize the results. Employing the general formalism developed in the papers \cite{BKL2,BKL4,BKL6}, we obtained the explicit expressions for the average number of twisted photons radiated by a relativistic charged particle moving in an elliptic undulator. To account for the quantum recoil, we used the Baier-Katkov method \cite{BaiKat1,BaKaStrbook} adapted in \cite{BKL4,BKL6} to describe the inclusive probability of radiation of twisted photons in the domain of parameters where the main part of radiation is concentrated. In this domain, we derived the explicit expressions for the inclusive probability of radiation of one twisted photon by  charged scalar and Dirac particles. In the limit of negligibly small quantum recoil and in the paraxial approximation, these expressions reproduce the average number of twisted photons created by the classical current of a point charge moving in an elliptical undulator. The last expression was also obtained in the present paper. In the particular cases of circular and planar undulators, the expressions for the probability of twisted photon radiation in elliptical undulators derived in this paper turn into the known ones \cite{BKL2,BKL4,BKL6}. It was proven that radiation of twisted photons obeys the selection rule: $m+n$ is an even number, where $m$ is a projection of the total angular momentum of a twisted photon and  $n$ is the undulator radiation harmonic number. This rule holds for both classical and quantum approaches for describing radiation. It is the generalization of previously known selection rules for radiation of twisted photons by circular and planar undulators and it is valid for the class of trajectories of the form \eqref{traj_odd}.

The explicit expressions for the one-particle amplitude of twisted photon radiation that were derived in the present paper allow one to find the average number of twisted photons emitted by the beam of charged particles moving in an elliptical undulator. To this end, one can employ the theory of radiation of twisted photons by particle bunches developed in \cite{BKb,BKL5,BKLb}. Since the motion of charged particles in the plane laser wave with an elliptical polarization is similar to the motion of charged particles in an elliptical undulator, the formulas derived in this paper can be used to describe radiation of twisted photons in both cases \cite{BKL4}. Moreover, the developed formalism can be employed for description of radiation of twisted photons in axial and planar channelling \cite{ABKT,EpJaZo}. The similar properties of radiated twisted photons are expected for transition radiation in dispersive medium with the permittivity tensor invariant under translations along the elliptical helix \cite{BKL5,choler_pert} and for the elliptic undulators filled with dispersive medium \cite{BKL8}.

\paragraph{Acknowledgements.}

We are indebted to P.~S. Korolev and G.~Yu. Lazarenko for useful comments. The work was supported by the RFBR grant No. 20-32-70023.

\appendix
\section{Proof of the selection rule}\label{Sel_Rul}

Consider the radiation of twisted photons by a charged particle moving along the trajectory \eqref{traj_odd}. Then, for the factor entering the integrand of $I_3$, we have
\begin{equation}
\begin{split}
    \exp\bigg(i\sum_{k=-\infty}^\infty k_3d_k e^{2ik\vf} \bigg)&=e^{ik_3d_0}\prod_{k=1}^\infty\exp\big[i(k_3d_k e^{2ik\vf}+k_3d^*_k e^{-2ik\vf} )\big]=\\
    &=e^{ik_3d_0}\sum_{\{k_r\}}\exp\bigg(i\sum_{r=1}^\infty 2rk_r\vf \bigg) \prod_{r=1}^\infty i^{k_r}j_{k_r}(2k_3d_r,2k_3d_r^*),
\end{split}
\end{equation}
where, in the last expression, the summation over all the integer-valued sequences $\{k_r\}$, $r=\overline{1,\infty}$, with a finite number of nonzero elements is understood. Analogously, using repeatedly the addition theorem \eqref{add_thm} and the property \eqref{j_prop}, the another factor in the integrand of $I_3$ can be written as
\begin{equation}
    j_{m}(k_\perp x_+,k_\perp x_-)=\sum_{\{l_q\}}\de_{m,\sum_q l_q}\exp\bigg(im\vf+i\sum_{q=-\infty}^\infty 2ql_q\vf \bigg) \prod_{q=-\infty}^\infty j_{l_q}(k_\perp c_q,k_\perp c_q^*),
\end{equation}
where the sum runs over all the integer-valued sequences $\{l_q\}$, $q=\overline{-\infty,\infty}$, with a finite number of nonzero elements. As a result, the integrand of $I_3$ becomes
\begin{equation}
\begin{split}
    \dot{x}_3j_m(k_\perp x_+,k_\perp x_-)&\exp\bigg(i\sum_{k=-\infty}^\infty k_3d_k e^{2ik\vf} \bigg)= e^{ik_3d_0}\sum_{\{k_r\},\{l_q\},n'} 2i\omega n'd_{n'} \de_{m,\sum_q l_q} e^{in\vf}\times\\
    &\times\prod_{r=1}^\infty \Big[i^{k_r}j_{k_r}(2k_3d_r,2k_3d_r^*)\Big] \prod_{q=-\infty}^\infty \Big[j_{l_q}(k_\perp c_q,k_\perp c_q^*)\Big],
\end{split}
\end{equation}
where $2i\omega n'd_{n'}$ should be replaced by $\be_3$ for $n'=0$ and the number of the undulator radiation harmonic is
\begin{equation}
    n=m+\sum_{r=1}^\infty 2rk_r+\sum_{q=-\infty}^\infty 2ql_q+2n'.
\end{equation}
It is evident that $m+n$ is an even number. The integrands of $I_\pm$ are expanded in the same way. These expansions also respect the relation that $m+n$ is an even number. Furthermore, following along the lines of Sec. \ref{Rad_Tw_Phot}, it is not difficult to verify that the quantum recoil does not violate this selection rule, at least, in the case when $P_0$ is constant.

\section{Some properties of the functions  $g_{nm}$}\label{Gnm_Prop}

The functions $g_{nm}$ possess the symmetry properties
\begin{equation}\label{g_prop}
    g_{nm}(\vk,\rho,\de)=(-1)^m g_{n,-m}(\vk,\de,\rho),\qquad g_{nm}(\vk,\rho,\de)=g_{-n,m}(-\vk,\rho,\de).
\end{equation}
The following recurrence relations hold:
\begin{equation}
\begin{split}
    2mg_{nm}&=\rho(g_{n+1,m+1}+g_{n-1,m-1})+\de(g_{n+1,m-1}+g_{n-1,m+1}),\\
    2\frac{\partial g_{nm}}{\partial\vk}&=g_{n-2,m}-g_{n+2,m},\\
    2\frac{\partial g_{nm}}{\partial\rho}&=g_{n-1,m-1}-g_{n+1,m+1},\\
    2\frac{\partial g_{nm}}{\partial\de}&=g_{n+1,m-1}-g_{n-1,m+1},
\end{split}
\end{equation}
where $g_{nm}\equiv g_{nm}(\vk,\rho,\de)$. The generating function is written as
\begin{equation}
\begin{split}
    \sum_{n,m=-\infty}^\infty &g_{nm}s^nt^m = \exp\Big[\frac{\vk}{2}\Big(s^2-\frac{1}{s^2}\Big) +\frac{\rho}{2}\Big(st-\frac{1}{st}\Big) +\frac{\de}{2}\Big(\frac{t}{s}-\frac{s}{t}\Big)\Big],\\
    g_{nm}&=\int_{-\pi}^\pi \frac{d\vf d\psi}{(2\pi)^2} \exp\Big\{i\big[\vk\sin(2\psi)+\rho\sin(\vf+\psi)+\de\sin(\vf-\psi)-n\psi-m\vf\big]\Big\}.
\end{split}
\end{equation}
Employing the generating function, it is not difficult to prove the sum rules
\begin{equation}
\begin{gathered}
    \sum_{m=-\infty}^\infty g_{nm}(\vk,\rho,\de)=J_n(\rho-\de,\vk),\qquad \sum_{n=-\infty}^\infty g_{nm}(\vk,\rho,\de)=J_m(\rho+\de),\\
    \sum_{n,m=-\infty}^\infty g_{nm}(\vk,\rho,\de)=1,
\end{gathered}
\end{equation}
where $J_n(x,y)$ is a generalized Bessel function of two arguments \cite{Bord.1,Diden79,NikRit64,Dattol90,Dattol91}.

\end{document}